\documentclass[preprint,iop,amsmath,amssymb]{revtex4-1}
\usepackage{bm}
\usepackage{epsfig}
\usepackage{amsmath}
\usepackage{color}
\newcommand{\nix}[1]{}
\usepackage[unicode=true,colorlinks=true,urlcolor=blue,citecolor=blue]{hyperref}
\usepackage{subcaption}

\begin{document}
	
	\title{Fingerprints of the electron skew-scattering on paramagnetic impurities in semiconductor systems}
		\author{M.A. Rakitskii}
	\email{mixrak@gmail.com}
	\affiliation{Ioffe Institute, 194021 St.Petersburg, Russia}
		\author{K.S. Denisov}
	\affiliation{Ioffe Institute, 194021 St.Petersburg, Russia}
		\author{I.V. Rozhansky}
	\affiliation{Ioffe Institute, 194021 St.Petersburg, Russia}
		\author{N.S. Averkiev}
	\affiliation{Ioffe Institute, 194021 St.Petersburg, Russia}
	
	\begin{abstract}
 
 In this paper we argue that the electron skew-scattering on paramagnetic impurities in non-magnetic systems, such as bulk semiconductors, possesses a remarkable fingerprint allowing to differentiate it directly from other microscopic mechanisms of the emergent Hall response. We demonstrate theoretically that the exchange interaction between the impurity magnetic moment and mobile electrons 
  leads to the emergence of an electric Hall current persisting even at zero electron spin polarization. We describe two microscopic mechanisms behind this effect, namely the exchange interaction assisted skew-scattering and the conversion of the SHE induced transverse spin current to the charge one owing to the difference between the spin-up and spin-down conductivities. We propose an essentially all-electric scheme based on a spin-injection ferromagnetic-semiconductor device which allows one to reveal the effect of paramagnetic impurities on the Hall phenomena via the detection of the spin polarization independent terms in the Hall voltage.

	\end{abstract}


	\date{\today}

	\newcommand{\p}{\partial}
	\renewcommand{\d}{\operatorname{d}}
	\newcommand{\bra}[1]{\langle #1 |}
	\newcommand{\ket}[1]{| #1 \rangle}
	\renewcommand{\Re}{\operatorname{Re}}
	\renewcommand{\Im}{\operatorname{Im}}
	\newcommand{\St}{\operatorname{St}}

	\maketitle










\textit{Introduction.}
Nowadays, the spin phenomena physics in solids is particularly focused on the effects arising from the combination of a spin-orbit interaction and magnetism. 
This interplay lies in the basis of the modern spintronics and has proved to be a source of profound physics as well as device applications~\cite{Dyakonov,sinova2012new,vzutic2004spintronics,bader2010spintronics}. 
One example is the physics of the anomalous Hall effect (AHE) which 
remains a vibrant topic since its discovery
more than a century ago. 
Not only it plays the key role for the electrical detection of the sample magnetization, 
but also it enriches our understanding of fundamental processes in solids~\cite{AHE-Sinova}. 
The modern viewpoint is that the Hall effect measurements constitute a particularly effective tool allowing to probe rather subtle effects connected with both the exchange and  spin-orbit interactions. 
In particular, the formation of magnetic skyrmions due to Dzyaloshinskii-Moriya interaction has been observed be means of the topological Hall effects~\cite{MnSiAPhase,raju2019evolution}. 
The combined effect of the mixed real-momentum space topology is predicted to influence the Hall resistivity~\cite{Lux2020,Berry1,ishizuka2018spin} of topological systems, such as the magnetic topological insulators~\cite{yasuda2016geometric,tokura2019magnetic,chang2013experimental}. 

Various scenarios of  AHE and the spin Hall effect (SHE) emergence have been widely discussed, often classified into intrinsic and extrinsic mechanisms~\cite{Dyakonov,AHE-Sinova}.
One of the particularly interesting  extrinsic   
mechanisms concerns the
skew scattering of electrons on paramagnetic impurities 
embedded in a non-magnetic host. 
In the mid-70th it was argued to be responsible for AHE in dilute magnetic alloys~\cite{fert1972left,fert1976skew,fert1981hall,giovannini1973skew}, 
where the skew scattering turns out to be strongly enhanced
by the resonant coupling of itinerant electrons with magnetic impurity states. 
Despite that AHE in typical ferromagnets is dominated by other mechanisms, the paramagnetic impurities preserve the key significance for AHE in non-magnetic materials.
For instance, AHE due to paramagnetic centers has been recently confirmed in non-magnetic ZnO-based heterostructures
 ~\cite{maryenko2017observation}.
Moreover, the skew scattering on paramagnetic impurities has lately experienced a new twist in view of the emergent spin Hall effect. 
This process is debated~\cite{guo2009enhanced} to be the key origin of 
the strong enhancement of SHE and the giant	spin Hall angle observed in gold~\cite{seki2008giant}
Despite the high importance of the 
skew-scattering on paramagnetic centers in the formation of 
both SHE and AHE in non-magnetic systems, 
no clear physical concept allowing for a direct experimental verification of this effect 
has been formulated so far.
In this paper we 
argue, that
 the electron scattering on paramagnetic centers
  has
 remarkable fingerprints
 and can be 
  differentiated 
from other mechanisms of SHE and AHE in an experiment.

Indeed,
the conventional
mechanisms
of AHE or SHE
are based on
 the charge carrier motion asymmetry 
  due to spin-orbit coupling (SOC), which is  
 sensitive to the carrier spin $\bm S$.
From the symmetry point of view, 
the skew-scattering leading to 
 SHE can be
described as 
the contribution to the scattering rate of the form  $ \bm{S} \cdot (\bm k \times \bm k')$, where ${\bm k},{\bm k'}$ are the incident and scattered electron wavevectors, respectively.
In fact, the product $(\bm k \times \bm k')$ responsible for the scattering asymmetry and Hall response must go along with some vector of magnetic moment to meet the rime-reversal symmetry requirement.
Here $\bm S$ belongs to the incident electrons and the skew-scattering probability has opposite asymmetry for spin-up and spin-down electrons leading to SHE.
The associated AHE results from 
conversion of the spin current to the charge current 
due to the 
spin polarization of the carriers in a magnetic system.
Following this mechanism
AHE 
can emerge 
only provided that the charge carriers possess a finite spin polarization;
in particular, this conclusion is
confirmed experimentally
for semiconductor systems~\cite{chazalviel1972experimental,cumings2006tunable}.
However, 
	 this physical picture becomes different 
	when 
	a scatterer possesses its own magnetic moment $\bm J$. 
Firstly, one can construct the similar contribution to the skew-scattering rate by substituting the electron spin by that of the impurity, i.e. $ \bm{J} \cdot (\bm k \times \bm k')$. 
The important difference from the previously considered case is that now the electron spin is decoupled from
the quantities defining the asymmetry of the scattering.
The fingerprint of this effect is that it leads to the same scattering asymmetry regardless the electron spin state, 
resulting 
in a finite AHE response even 
for completely non-polarized electron gas.
Secondly, the presence of an impurity magnetic moment $\bm{J}$ renormalizes the transport scattering times for the spin-up and spin-down electrons 
leading to $\tau_{\uparrow} \neq \tau_{\downarrow}$ and, consequently, to the 
spin dependence of the 
 conductivity.
	As a result, the skew-scattering 
	induced by the 
	common SOC-term $ \bm{S} \cdot (\bm k \times \bm k')$ 
	leads 
	not only to SHE, but also induces the 
	transverse charge current 
already for non-polarized electrons.
Hence, if the magnetic impurities play a role in the observed Hall response, 
the latter should contain a contribution, which is independent of the electron spin polarization. 
Finding signatures of the AHE which
remain unchanged upon varying 
carrier spins would be a direct confirmation of this effect. 
In this paper we theoretically demonstrate that paramagnetic impurities in typical zinc-blend semiconductors (ZBS) indeed give rise to 
AHE 
independent of the electron spin polarization. 
We argue that the two microscopic mechanisms underlying this effect, namely the $ \bm{J} \cdot (\bm k \times \bm k')$ skew-scattering and 
{the combination of SHE and the spin dependence of the   conductivity}
have 
the same order of magnitude and  should be treated on the equal footing. 
{
Moreover, these mechanisms appear to be differently coupled to the $s-d$ and $p-d$ types of the exchange interaction.
We also discuss how the specific signatures of 
the skew-scattering on paramagnetic centers 
can be revealed experimentally.
We suggest to differentiate it from other AHE mechanisms using a 
device scheme with the spin injection from a ferromagnet into a semiconductor.
Alternatively, the spin injection into the semiconductor can be done by means of 
optical orientation. 

\textit{Microscopic consideration.}
Let us consider the scattering of an electron from conduction band states described by $\Gamma_6$ representation. 
The diagonal matrix elements for the core SOC term in the conduction band are zero, hence,  
for the $\Gamma_6$ band the effect of the SOC appears due to an admixture of the SOC split valence band states $\Gamma_7$ and $\Gamma_8$ via $\bm{k \cdot p}$ term. 
The conductance band wavefunction with account for the admixture of the valence band has the form~\cite{ivchenko2012superlattices,abakumov1972anomalous}:
\begin{align}
\label{eq_WF-ZBS}
& \Psi_{\bm k s} = e^{i \bm k \bm r} \Bigl( S + i \bm R (A \bm k - iB \left[\hat{\bm \sigma} \times \bm k \right] ) \Bigr) \ket{\chi_s},
\notag
\\
& A = P \frac{3E_g + 2 \Delta}{3E_g(E_g + \Delta)},
\quad
B = -P \frac{\Delta}{3E_g(E_g + \Delta)},
\end{align}
where $S$ and $\bm R = ({X}, {Y}, {Z})$
are the conduction and the valence band Bloch amplitudes at the
$\Gamma$--point, 
$\ket{\chi_s} = |{\uparrow,\downarrow}\rangle$ denote the electron spin state,
$\hat{\bm \sigma}$ is the vector of Pauli matrices,
$E_g$ is the band gap between $\Gamma_6$ and $\Gamma_8$ bands,
the real parameter corresponding to the momentum matrix element is $P =\left( i \hbar/{m_0} \right)\bra{S} p_x \ket{X}$.
The core SOC is taken into account via
the energy splitting $\Delta$
between $\Gamma_7$ and $\Gamma_8$ subbands.
We model the paramagnetic impurity by
a short-range potential acting
on electrons both by means of the electrostatic $u(\bm{r})$  
and the exchange 
interactions~\cite{giovannini1973skew}:
\begin{equation}
\label{eq_IMP1}
\hat{V}(\bm r) = u(\bm{r}) \hat{I} + \hat{u}_X(\bm{r})  \bm J \cdot \hat{\bm \sigma}.
\end{equation}
The second term describes the effective exchange interaction operator $\hat{u}_X$ of an itinerant electron and the impurity d-shell electrons. Since this term is essentially local 
one should distinguish the matrix elements of  $\hat{u}_X$ for $s-d$ and $p-d$ exchange interaction types~\cite{GajKos}.
In what follows we assume that some external magnetic field is applied to polarize paramagnetic impurities along $z$-direction. Given this condition we ignore the possible spin-flip processes for the impurities 
assuming $\bm J = J \bm e_z$ to be a constant vector 
directed along $z$ axis.
The scattering rate
i.e. the number of transitions 
from $(\bm k', s')$ to $(\bm k, s)$ state 
per second is  
determined 
 by the square modulus of the scattering $T$-matrix.
To account for the skew scattering in the leading order in the scattering potential it can be presented in the form~\cite{sinitsyn2007semiclassical}:
\begin{align}
\label{eq-T2}
&
| T_{kk'}^{ss'} |^2 \approx
 |V_{kk'}^{ss'}|^2 + W_{kk'}^{ss'}, \\
&
W_{kk'}^{ss'} =
2\pi g_0 \sum_{\mu=\uparrow,\downarrow}
\left\langle
\Im (V_{k'k}^{s's} V_{kq}^{s \mu} V_{q k'}^{\mu s'})
\right\rangle_q,
\notag
\end{align}
where $V_{kk'}^{ss'}$ is the matrix element of the impurity potential, and
$W_{kk'}^{ss'} = - W_{k'k}^{s's}$ is an asymmetric term responsible for the skew-scattering, here 
$g_0$ is the electron density of states at the Fermi~level, 
$\langle \dots \rangle_q$ means
averaging over the intermediate momentum $\bm q$ directions.

The matrix element $V_{kk'}^{ss'}$ of the potential (\ref{eq_IMP1}) 
between $(\bm k', s')$ and $(\bm k, s)$ states includes 
the skew-scattering relevant contributions $(\bm{k}\times \bm{k}')$:
\begin{align}
\label{eq_ME}
& 
\hat{V}_{kk'} = u_0 \hat{U}_{kk'} + \hat{M}_{kk'},
\\
& \hat{U}_{kk'} = 1 + i(2AB + B^2) (\bm k \times \bm{k'}) \cdot \hat{\bm \sigma},
\notag
\\
& \hat{M}_{kk'} =  
\alpha_{ex} \bm J \cdot \hat{\bm \sigma} + i \beta_{ex} (2AB - B^2) \left( [\bm k \times \bm{k'}] \cdot \bm J \right). 
\notag
\end{align}
Here we assumed that the matrix element of the electrostatic (spin-independent) part of the impurity potential $u_0 = \bra{S \bm{k}} u \ket{S \bm{k}'} = \bra{X \bm{k}} u \ket{X \bm{k}'}$ does not depend on either the electron momentum ($u(\bm{r})$ is localized within the Fermi wavelength) or the Bloch amplitude ($u(\bm{r})$ changes slowly within the unit cell)~\cite{abakumov1972anomalous}. 
The exchange interaction, on the contrary, is described by two parameters $\alpha_{ex} = \bra{S} \hat{u}_X \ket{S},\beta_{ex} = \bra{X} \hat{u}_X \ket{X}$ corresponding to the $s-d$ and $p-d$ exchange coupling, respectively~\cite{GajKos}.  
 We assume that the electrostatic part
significantly exceeds the exchange one $u_0\gg u_X$.
Keeping only the leading terms with respect to $u_X/u_0$ we find that the skew-scattering contribution to the scattering rates $W_{kk'}^{ss'}$ includes two terms:
\begin{align}
\label{eq_skewRateZBs}
& W_{kk'}^{ss'} =
- 2 \pi g_0 u_0^2
\left[ \bm k \times \bm{k'} \right]_z
\left(
\hat{\sigma}_{ss'}^z Z_{0} +
\delta_{ss'} Z_{X}
\right), 
\notag
\\
& Z_{0} = u_0 \left( 2AB + B^2 \right),
\quad
Z_{X} = J\beta_{ex} \left(2 AB - B^2 \right) + 2 J \alpha_{ex} \left( 2AB + B^2 \right). 
\end{align}
The contribution $Z_0$ is not aware of the impurity magnetic moment, it is therefore responsible for the spin-dependent skew-scattering, described 
as $\bm{S} \cdot (\bm k \times \bm k')$.
The contribution $Z_X$, on the other hand, exists only if the impurity spin is taken into account and describes the  spin-independent skew-scattering $ \bm{J} \cdot (\bm k \times \bm k')$.
Note, that the $\bm{J}$-dependent  skew-scattering appears in the first order in 
$u_X/u_0$. 
It is important to emphasize, that in this order there exists another effect eventually leading to the similar spin-independent Hall response.
The first Born 
approximation term in the $T$-matrix, which determines the transport scattering time 
appears to be spin-dependent (see Eqs.~\ref{eq-T2},\ref{eq_ME}).
Consequently, 
the scattering times $\hbar/2\tau_{\uparrow,\downarrow} = n_i \pi g_0 (u_0 \pm J \alpha_{ex})^2 \approx (\hbar/{2\tau}) \cdot (1 \pm 2 J \alpha_{ex}/u_0)$ differ,
here we defined the average scattering time as $\hbar /2\tau = n_i \pi g_0 u_0^2$, and $n_i$ is the impurity concentration. 
The difference between $\tau_{\uparrow,\downarrow}$ is in the first order in~$u_X/u_0$ appears to be 
\begin{equation}
\label{eq_DeltaTau}
 \Delta \tau \equiv | \tau_\uparrow - \tau_\downarrow | \approx \tau (4 \alpha_{ex} J/u_0).
\end{equation}
Due to this difference 
a spin Hall current 
due to the non-magnetic $\bm{S} \cdot (\bm k \times \bm k')$ skew-scattering ($Z_0$ contribution) is converted into transverse electrical current.

\textit{Anomalous Hall effect.}
Let us now proceed with considering the Hall conductivity 
of the non-magnetic ZBS with degenerate electron gas doped by paramagnetic impurities 
described by Eq.~\ref{eq_IMP1}.
We assume that the impurity concentration $n_i$ is low enough 
so that the conductance band structure is not affected by the exchange part of the impurity 
and the electrons remain non-polarized.  
Since the asymmetrical part of the scattering cross-section leading to the skew-scattering 
is typically small compared 
to the total cross-section, 
the longitudinal conductivity is 
given by 
$\sigma_{0} = \sigma_{xx}^\uparrow + \sigma_{xx}^\downarrow$, 
where $\sigma_{xx}^{\uparrow,\downarrow} = n_{\uparrow, \downarrow} e^2 \tau_{\uparrow, \downarrow}/ m$ 
correspond to the spin-up and spin-down electron conductivities with
the transport scattering 
times $\tau_{\uparrow,\downarrow}$; 
the total electron concentration is $n = n_\uparrow + n_\downarrow$. 
The associated Hall conductivity 
for each spin channel 
$\sigma_H^{\uparrow,\downarrow}\equiv
\sigma_{yx}^{\uparrow,\downarrow}$ can be expressed 
 in terms of 
the Hall angle $\sigma_H^{\uparrow,\downarrow} = \theta_H^{\uparrow,\downarrow} \sigma_{xx}^{\uparrow,\downarrow}$. 
We calculated the Hall angles on the basis of the Boltzmann kinetic equation and obtained the following expressions:
\begin{equation}
\label{eq-Hangle1}
\theta_H^{\uparrow,\downarrow} = 
\left(\Omega_{X} \pm \Omega_{0} \right) \tau_{\uparrow,\downarrow}, \qquad
\Omega_{0,X} =  Z_{0,X} \cdot
\frac{2 \pi}{3 \tau} g_0 k_F^2,
\end{equation}
where 
$k_F$ is the Fermi wave-vector, and 
$\Omega_{0,X}$ are the effective cyclotron frequencies, which account for the  $\bm{S} \cdot (\bm k \times \bm k')$  and $\bm{J} \cdot (\bm k \times \bm k')$ skew-scattering 
terms 
respectively. 
The total anomalous Hall conductivity $\sigma_H^A$ in the first order in $u_X/u_0$ can be expressed in the following way: 
\begin{equation}
\label{eq_SigAHE}
\sigma_H^A = \sigma_{H}^\uparrow + \sigma_H^\downarrow
= \sigma_{0} \left( P_s \cdot \theta_0 + \theta_X -  \theta_0 \cdot \frac{\Delta \tau}{\tau}  \right),
\qquad
\theta_{0,X} \equiv \Omega_{0,X} \tau, 
\end{equation}
where $\sigma_0 = n e^2 \tau /m \approx \sigma_{xx}$ is the total longitudinal conductivity, 
and 
$P_s = (n_\uparrow - n_\downarrow)/ n $ 
is the spin polarization 
of the electron gas, $\Delta \tau$ is defined by Eq.~\ref{eq_DeltaTau}. 
\begin{figure}
	\begin{subfigure}{0.45\textwidth}
		\begin{minipage}{0.1\textwidth}
			\vbox{\caption{} \vspace{3cm}}
		\end{minipage}
		\begin{minipage}{0.85\textwidth}
			\includegraphics[width=\textwidth]{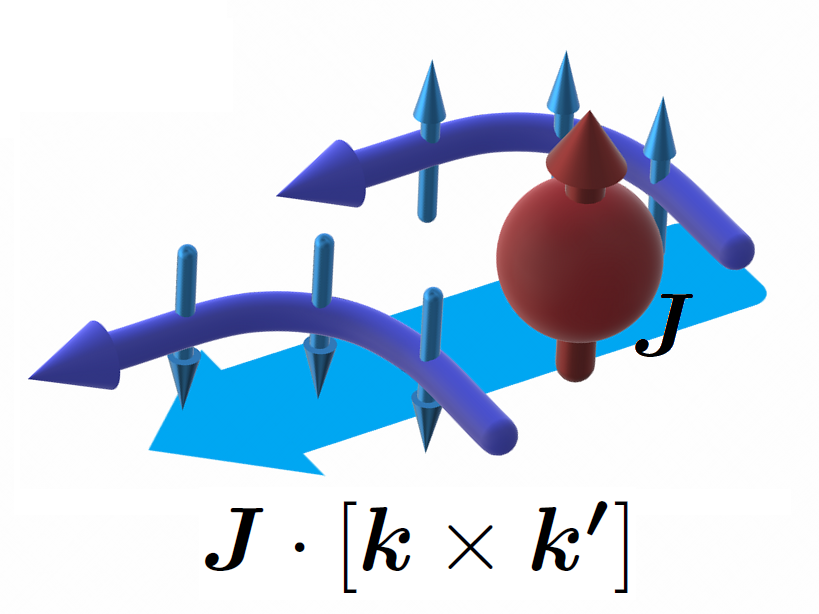}
		\end{minipage}
	\end{subfigure}
	\begin{subfigure}{0.45\textwidth}
		\begin{minipage}{0.1\textwidth}
			\vbox{\caption{} \vspace{3cm}}
		\end{minipage}
		\begin{minipage}{0.85\textwidth}
			\includegraphics[width=\textwidth]{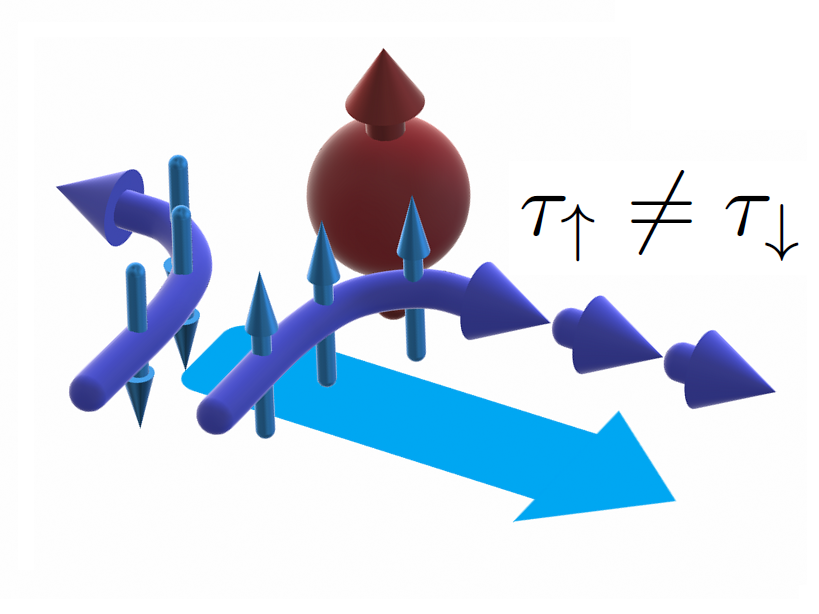}
		\end{minipage}
	\end{subfigure}
	\caption{Two mechanisms of the Hall effect due to skew scattering on a paramagnetic center.}
		\label{current1}
\end{figure}
Here the first contribution to $\sigma_H^A$ is
proportional to 
the spin polarization $P_s$ as 
it is associated with the spin-dependent skew-scattering ($Z_0$) and
spin to charge transverse current conversion.
In non-magnetic systems the equilibrium
concentrations of electrons in two spin subbands
are equal $n_\uparrow = n_\downarrow$ 
and this contribution vanishes. 
The other two 
contributions are sensitive to the $\bm{J}$-dependent scattering.
They contribute to the AHE provided the paramagnetic impurities are  polarized by an external magnetic field. 
The second term $\sigma_0 \theta_X$ 
arises from $\bm{J} \cdot (\bm k \times \bm k')$ skew-scattering (Fig.~\ref{current1} (a)), which is 
unaware 
of the electron spin state thus leading to  
AHE even for unpolarized carriers. 
The third term stems from the combination of SHE scattering described by the Hall angle $\theta_0$ and the $\bm{J}$-dependent difference  between the transport scattering times $\Delta \tau$. 
It can be interpreted as follows: 
as soon as the $\bm{J}$-dependent scattering makes 
the longitudinal currents of the spin-up and spin-down electrons different $\sigma_{xx}^\uparrow \neq \sigma_{xx}^\downarrow$, 
the corresponding transverse currents arising from 
$\bm{S} \cdot \left[\bm{k}\times \bm{k}'\right]$ skew-scattering and 
flowing in the opposite direction for spin-up and spin-down electrons become also different by their magnitude,
thus restoring the transverse electric current, as schematically illustrated in  Fig.~\ref{current1} (b).
Essentially, 
these mechanisms contribute with the same order of magnitude to~$\sigma_H^A$:
\begin{equation}
{\sigma_H^A} = \sigma_0\left( \theta_X -  \theta_0 \frac{\Delta \tau}{\tau_0} \right) = 
\sigma_0 \cdot  J \Bigl[ \beta_{ex} \left(2AB - B^2\right) - 2 \alpha_{ex} \left(2 AB + B^2 \right) \Bigr] \cdot \frac{2\pi}{3} g_0 k_F^2, 
\end{equation}
so they 
are equally important for producing AHE in  a non-magnetic media. 
Note, however, that the spin dependence of the conductivity is sensitive only to the $s-d$ exchange interaction constant  $\alpha_{ex}$, while the $\bm{J}$-dependent skew-scattering is also affected by $p-d$ coupling~$\beta_{ex}$.
The theoretical considerations given above naturally suggest that contribution to AHE driven by paramagnetic impurities 
should exist even for zero electron spin polarization and exhibit the magnetic-field dependence 
following 
the Brillouin function for the impurity magnetic moments, similarly to the recently observed AHE in ZnO-systems~\cite{maryenko2017observation}. 

\begin{figure}
	\begin{subfigure}{0.45\textwidth}
		\begin{minipage}{0.1\textwidth}
			\vbox{\caption{} \vspace{3cm}}
		\end{minipage}
		\begin{minipage}{0.85\textwidth}
			\includegraphics[width=\textwidth]{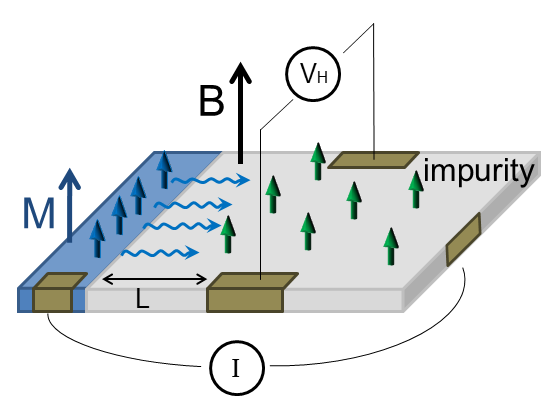}
		\end{minipage}
	\end{subfigure}
	\begin{subfigure}{0.45\textwidth}
		\begin{minipage}{0.1\textwidth}
			\vbox{\caption{} \vspace{3cm}}
		\end{minipage}
		\begin{minipage}{0.85\textwidth}
			\includegraphics[width=\textwidth]{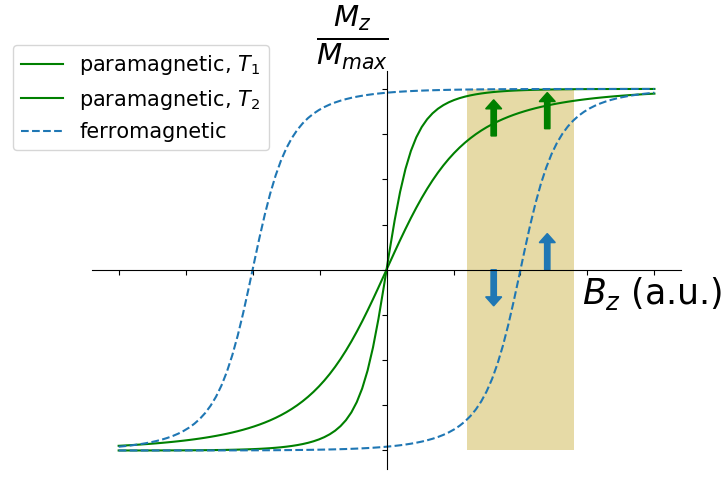}
		\end{minipage}
	\end{subfigure}
	\caption{Illustration of an experiment with a spin injection from FM to ZBS (a), polarization dependence of the magnetic dopant and injected electrons on an external magnetic field (b) }	
	\label{f2}
\end{figure}

\textit{Setting with nonequilibrium spins.}
The 
skew-scattering on paramagnetic impurities
can be directly differentiated from other spin-dependent mechanisms of the Hall response when operating with nonequilibrium spins.
Here we suggest 
an experimental setting shown in Fig.~\ref{f2} (a),
 which consists of a ferromagnetic injector in contact with ZBS doped with paramagnetic impurities.
The Hall contacts are placed at the distance $L$ away from the injector.
We assume that a 
longitudinal current $j_{x}$ is injected from FM 
and flows 
 across the semiconductor layer.
This current carries non-equilibrium spin polarization $P_s$, which is maintained due to the difference in the quasichemical potentials of the injected spin-up and spin-down electrons
$\mu_s \equiv (\mu_\uparrow - \mu_\downarrow)/2$.
The spin polarization
is partially lost 
as the injected electrons reach the Hall contacts: 
$P_s(L) \propto \mu_s(0) e^{-L/L_s}$, here $L_s$ is the spin diffusion length, and
$\mu_s(0)$ is the 
spin accumulation at 
the interface between FM and ZBS.
According to the spin injection theory~\cite{fabian2007semiconductor},
$\mu_s(0)$ 
is proportional to the total electric current flowing across the interface $\mu_s(0) =  - j_{x} \alpha_s L_s \sigma_0^{-1}$ where $\sigma_0$ is the ZBS conductivity 
and $\alpha_s$ is a material parameter describing the efficiency of spin injection through the interface.
The total longitudinal current 
$j_{x} = j_{x}^\uparrow + j_{x}^\downarrow$ 
$= \sigma_{xx}^{\uparrow} \nabla \mu_{\uparrow} + \sigma_{xx}^{\downarrow} \nabla \mu_{\downarrow}$, 
}
is accompanied by the appearance of the Hall currents carried by the spin-up and spin-down electrons
$j_H^{\uparrow,\downarrow} = j_{x}^{\uparrow,\downarrow}
\theta_H^{\uparrow,\downarrow}
$, here $\theta_H^{\uparrow,\downarrow}$ are
the corresponding 
Hall angles, see Eq.~\ref{eq-Hangle1}.
The condition of zero Hall electric current 
$j_\uparrow^H + j_\downarrow^H+\sigma_{xx}E_H=0$ 
leads to appearance of the transverse electric field $E_H \equiv \rho_H j_{x}$, 
where for the Hall resistivity $\rho_H$ in the lowest order in $u_X/u_0$ we obtain:
\begin{equation}
\rho_H = \rho_0 \left(
\theta_0 \cdot \alpha_s e^{-L/L_s} +
\theta_X - \theta_0 \frac{\Delta \tau}{\tau} 
\right), 
\qquad
\rho_0 = \sigma_0^{-1}.
\end{equation}
Here the first term stems from the conversion of spin current 
to electrical current 
due to 
the 
nonequilibrium spin polarization. 
The second and third terms 
 reveal 
  the presence of magnetic impurity induced skew-scattering 
and does not depend on spin  polarization.

Let us see how the two contributions to the Hall voltage can be separated using such a spin injection-based device. 
By applying 
a perpendicular magnetic field as shown in Fig.~\ref{f2} (a) the impurities get polarized.
Also, 
the external magnetic field controls magnetization of the ferromagnet injector and, hence, the injected electrons. 
Note, that due to the hysteresis of the ferromagnet there is a range of the magnetic fields where 
the spins of the impurities remain oriented in the fixed direction while the magnetization of the ferromagnet gets flipped, see Fig.~\ref{f2} (b).
The flip of the magnetization reverses 
the sign of the injected electrons spin polarization $\alpha_s \to - \alpha_s$ and thus inverts $\theta_0$ driven contribution to the Hall voltage.
Taking the difference between
the Hall voltages at the two sides of the magnetization reversal point ($\rho_H^{<}, \rho_H^{>}$) one can directly extract the contribution of the paramagnetic impurities in terms of the Hall angle:
\begin{equation}
	\label{eqdiff}
\theta_{H} \equiv
\frac{\sigma_H^A}{\sigma_0} = \frac{\rho_H^< - \rho_H^>}{2 \rho_0}.
\end{equation}
A non-zero quantity (\ref{eqdiff})
would be a direct manifestation of scatterers with intrinsic spin contributing to the Hall transport. 

\textit{Discussion and Summary.}
Let us estimate the magnitude of the Hall angle $\theta_H = \sigma_H^A/\sigma_0$ for typical ZBS due to zero spin polarization contributions to the AHE. 
We take the Fermi energy $E_F = 100$~meV and get $\theta_H \approx 3 \cdot 10^{-6} $ for GaAs doped by Mn ($\Delta = 0.346$~eV, $N_0 \alpha_{ex} = 0.2$~eV, $N_0 \beta_{ex} = -1.2$~eV ~\cite{Dietl2014DiluteFS})
and $\theta_H \approx 2 \cdot 10^{-5}$ for CdTe also with Mn impurity ($\Delta = 0.95$~eV, $N_0 \alpha_{ex} = 0.22$~eV, $N_0 \beta_{ex} = -0.88$~eV ~\cite{GajKos}). The obtained value is about an order of magnitude smaller than the typical magnitude of the AHE angles observed in various semiconductors~\cite{AronzonRozh,nagaosa2010ahe,cumings2006tunable}. 
The Hall angle magnitude can be further increased for systems with smaller band gap, such as InAs or InSb, or due to a resonant state at the paramagnetic dopant. 
In II-VI dilute magnetic systems the developed theory can be applied to the electron skew-scattering on bound magnetic polarons. In this case the effective polaron spin $J$ can be up to three orders of magnitude larger leading to a strong enhancement of the spin-independent contributions to AHE.

In summary, 
we have demonstrated that the combined action of SOC for the electrons in the conductance band and the exchange interaction with paramagnetic scatterers lead to the emergence of the electrical Hall current even for zero electron spin polarization. 
We have described two microscopic mechanisms behind this effect, namely the exchange interaction assisted skew-scattering and the conversion of the transverse spin current into the electrical one due to spin dependence of the scattering time.
The proposed effect can be 
directly checked in experiments by observing the corresponding spin-independent contribution to the Hall voltage. 

\textit{Acknowledgements.} 
Authors acknowledge the financial support from 
from Russian Science Foundation project $17-12-01182$ (theoretical model) and from 
Russian Foundation for Basic Research (RFBR) Grant No. $18-02-00668$.
K.S.D., I.V.R. and M.A.R thank the Foundation
for the Advancement of Theoretical Physics and Mathematics
"BASIS".

	\bibliography{Ref}

\end{document}